\documentclass[aps,twocolumn]{revtex4}%
\usepackage{amsfonts}
\usepackage{amsmath}
\usepackage{amssymb}
\usepackage{graphicx}%
\begin{document}
\title{\Large Complexity, time and music}
\author{Jean Pierre Boon}
\email{jpboon@ulb.ac.be}
\homepage{http://poseidon.ulb.ac.be}
\affiliation{Physics Department CP 231, Universit\'e Libre de Bruxelles, 1050 - Bruxelles, Belgium}

\date{\today}
\maketitle

\section*{ \large Time and complexity}

The formulation of the concept of {\em complexity} can be traced back to 
the late 19th century when the Austrian physicist Ludwig Boltzmann studied
the evolution of physical systems. Boltzmann gave a microscopic content to the 
thermodynamic {\em entropy} in connection with the irreversible time evolution 
towards equilibrium where systems exhibit maximum disorder or complexity.
It was another Austrian physicist, Erwin Schroedinger (also born in Vienna) 
who in the mid 20th century suggested that under non-equilibrium conditions, 
temporal evolution could produce ordered states with less complexity than the
equilibrium state, or rather with a different kind of complexity. 
Time then acquires a new status: it is  an evolutionary factor - 
{\em the arrow of time} - by which natural phenomena can 
make systems become organized, which organization materializes in 
the {\em emergence} of space- and time-dependent structures with various
degrees of complexity. The concept of emergence was at the core of new
insights that were highlighted in a 1972  article entitled
{\em More is different} [1] in which physicist P.W. Anderson emphasized that 
an ensemble of simple elements is more than merely their sum, and that the
the whole can exhibit remarkable properties that cannot be predicted from the
(even exact) knowledge of the constituting elements. Complexity appears as
a characteristic of the ensemble of these elements, which, in physics, are often 
called the {\em degrees of freedom} of the system.

Complexity in art pieces may be viewed as the materialization 
of the many degrees of freedom involved in artistic creation, and 
this is probably one of the many reasons why art resists theoretical analysis, 
theory being understood as a set of principles and methods from which logical 
computation and analysis can be performed on sets of data obtained 
from measurements carried on a system (here the piece of art). 
Considered from this point of view, music may be more easily
amenable to scientific analysis because of its underlying mathematical structure
and because, when performed, it steers a one-dimensional course, the
dimension of time.

The {\em arrow of time} is intrinsic to musical expression: music emerges from
silence and returns to silence. And, in contrast with other forms of art, it is not
possible to take a snapshot of a piece of music; if time were to be stopped, the
music simply vanishes. In practical terms, a musical  sequence can be considered 
as the time evolution of an acoustic signal, and therefore a piece 
of music can be cast in the form of a time series. Technically this means that,
be it in the form of a sequence of acoustic pulses or in the written form of symbols
in a music score, a piece of music can be cast as a set of data points distributed 
along an axis with the dimension of time. Such a set, however complicated it may be, can always be coded as a {\em string of bits}. 

This leads us to the more formal concept of complexity as
formulated in 1965 independently by G.J. Chaitin and A.N. Kolmogorov who
proposed an algorithmic (objective) definition. 
The idea goes as follows: given a string of bits, which
is the shortest computer program that is able to produce the string? It is clear that
a finite sequence of 2N bits with alternatively 1 and 0 (10101010101010101...) should 
be produced by a program with minimal length (write N times 1 followed by 0)
whereas a sequence with the same number of 1's and 0's distributed randomly (110100000111010100001011111...) can probably not be reproduced otherwise 
than by rewriting the complete string. The length of the program is then used to
quantify the degree of complexity. An interesting aspect to the procedure is that
one must - or the pointer of the computing machine must - scan sequentially the bits 
of the string in order to perform the computation, which is an operation performed 
in time. This observation indicates that the algorithmic measure of complexity
implies a measurement in time. In particular, considering the string of data
obtained from the coding of a piece of music, the algorithmic measure of its degree 
of complexity then yields a signature of the dynamics of that piece of music. And 
reciprocally the dynamics perceived in the music reflects - at least one of the
components of - its complexity.

\section*{\large  Complexity and music}

Tools developed in the context of dynamical systems theory in mathematics and
in physics provide techniques to analyze sets of data obtained from measurements on
physical and other systems which exhibit complex behavior and are very difficult 
to explain on the basis of classical causal laws. For instance the laws of gravitational motion explain beautifully the orbital movement of planets in the solar system, but
the energy dissipated in the erratic solar flares, turbulence in the atmosphere, 
cardiac arrhythmias, or stock market fluctuations cannot be described analytically by
first principle laws. Yet data analysis of the time series obtained from measurements
in such types of phenomena can provide quantitative measures from which
their degree of complexity can be evaluated, and thereby provide insights to the
mechanisms of complex behaviors. 

How can we apply these concepts to the analysis of music, and 
what insights would this approach provide as to our perception of music? 
The answer to the first question is somewhat technical [2]. 
One can take the printed score of a chosen piece of music and play it
on a synthesizer interfaced to a computer. The pitch values are converted 
into digital data which are stored in the computer memory where the score 
is now converted into a time series, say $X(t)$ for a single part score. 
Pieces with several parts are treated part by part to produce 
a set of time series $X(t), Y(t), Z(t),...$ .  Alternatively on can access 
data libraries where digitally coded music scores are readily available.
Once in the form of time series, data are processed with the tools developed 
in the context of dynamical systems theory.

\section*{\large  From time to space}

One particularly interesting characterization is obtained with the 
construction of the {\em phase portrait}: $X(t), Y(t), Z(t),...$ 
(or $X(t), X(t+n\Delta t), X(t+m\Delta t)$, using the time-delay method for
single part pieces).  Suppose we want to analyze a string trio piece, and we
processed the violin part, $X(t)$, the viola part, $Y(t)$, and the cello part, $Z(t)$.
We then construct a three-dimensional Euclidean space spanned by $X(t), Y(t), Z(t)$
called the {\em phase space}. The $X$-axis represents the range over which the
notes are played on the violin, and similarly for the viola along the $Y$-axis, and for
the cello along the $Z$-axis. Suppose the piece starts with the violin playing A, the
the viola playing F, and the cello playing C simultaneously on the first beat of the
first bar. Plotting the corresponding numerical values along each axis gives one
point in the $XYZ$ phase-space. The next note will give the next point in 
phase-space, and so on until the completion of the piece. Joining the points
yields a trajectory as illustrated in Fig.1c which shows the phase portrait of the 
three part {\it Ricercar} of Bach's {\it Musical Offering}.  
The result is called the {\em phase portrait} which gives a spatial
representation (in the abstract phase space) of the temporal dynamics of the
music piece reconstructed from the time series obtained from the pitch
variations as a function of time: the phase portrait maps the {\em time} evolution of a dynamical process onto a {\em space} representation.

\begin{figure}
% \resizebox{4cm}{!}{
% \twoimages{{Fig_4a}}{Fig_5}}
\begin{center}
\makebox{
\resizebox{5.5cm}{5.5cm}{
\includegraphics{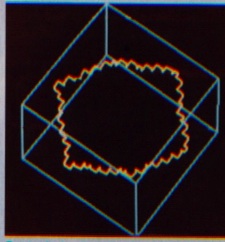}}}\\
\makebox{
\resizebox{5.5cm}{5.5cm}{ 
\includegraphics{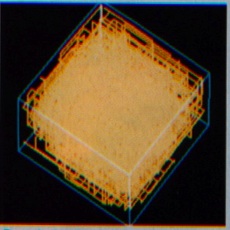}}}\\
\makebox{
\resizebox{5.5cm}{5.5cm}{
\includegraphics{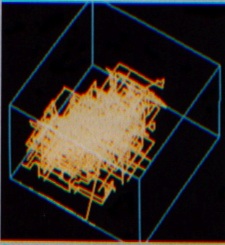}}}
\caption{Phase portraits: Ascending and descending chromatic scale (top) $D_f=1$; 
Computer generated random music (center) $D_f=3$; Ricercar from Bach's Musical 
Offering (bottom) $D_f=1.72$.}
\end{center}
\end{figure}

So measuring characteristics of the spatial object provides a measure of the 
dynamics. In figure 1, along with the phase portrait of the Bach's piece, two
typical extreme examples: an elementary score constructed as a canon of 
three repeatedly ascending and descending chromatic scales and
a piece of random music constructed with a computer generated white noise 
algorithm. Obviously the three pieces exhibit very different space occupations.
Dimensionality measures were also used in the context of plastic art: in the
two-dimensional space of Pollock's abstract paintings, an identification of 
Pollock's style is obtained by measuring the fractional value of the space covered
by the paint [3]. The same type of dimensionality analysis is performed for
the phase portraits in music with the box counting method. 
To illustrate the procedure we consider the pictures shown in figure 1. 
The repeated ascending and descending chromatic scales
are periodic in time so that their corresponding trajectory in space form a closed loop
which has dimensionality $D_f=1$. In contrast, a piece of random music explores
all possible combinations in the course of its time evolution and therefore the resulting trajectory fills homogeneously the entire available space: the resulting object has dimensionality $D_f=3$. 
On the one hand, we have an almost totally predictable musical piece, the
chromatic scales with minimal complexity, and on the other hand a sequence 
of unpredictable subsequent sounds, therefore with maximal complexity. 
The piece of Bach lies in the intermediate range with a dimensionality $D_f=1.72$.
The dimension $D_f$ (known as the Hausdorff dimension) computed by the box-counting method, characterizes the structure of the complete phase trajectory, and its value yields a quantitative evaluation of the global dynamics of the music piece.

\section*{\large  Global dynamics and local dynamics}

While the Hausdorff dimension provides an evaluation of the {\em global} dynamics 
of a piece of music, a measure of the {\it local} dynamics can be obtained from the application of information theory [2]. The analysis proceeds on the basis of the data files
and  goes along the lines of the discussion in the introductory section.
A sequence of notes is viewed as a string of characters and is analyzed from the 
point of view of its information content. The string is defined by straightforward 
coding of the pitch by assigning a symbol to each note. 
The Shanon {\em entropy} $H$ measures the information content of a string of characters on the basis of their occurrence probability. It is defined such that its value has an upper bound ($=1$) for a fully random sequence. $H_0$ is the zeroth order entropy which is a measure of the straight occurrence of each note, and the $\alpha$-th order entropies, 
$H_\alpha$ ($\alpha \neq 0$), follow from the successive conditional 
probabilities at increasing orders. In fact the most relevant quantity is $H_{\alpha=1}$
which is the probability to find the note $s_{i+1}$ given that the previous note 
was $s_i$. In addition, in western music, an important  feature that must be accounted for is {\em tonality} (here noted $\theta$). One therefore introduces a quantity defined as the 
{\em parametric entropy} $H'_1$ which measures 
the information content of a musical sequence quantifying the transition 
probabilities from one note to the next given that the transition can occur 
within the reference tonality ($(s_i,s_{i+1}) \in \theta$), outside the tonality
($(s_i,s_{i+1}) \not \in \theta$), or from $\theta$ to off  $\theta$ 
($s_i\in \theta,s_{i+1}\not \in \theta$), and vice-versa. 
The operational result is that a large value of the parametric
entropy is indicative of frequent excursions away from the tonality, with
transitions over intervals distributed over a large number of notes. On
the contrary, the parametric entropy has a low value when a note determines
almost unambiguously the next one, in particular when the next note remains
in the range of tonality.  
 
Dimensionality and entropic analyses performed on eighty sequences chosen in 
the music literature from the 17th century (J.S. Bach) to the 20th century (E. Carter)
lead to interesting observations. When the values of $D_f$ and $H'_1$ are
organized in chronological order (referring to the date of composition) - 
with very few exceptions -  there is no obvious clustering of pieces by 
composer or by period of composition; this holds for dimensionality as well 
as for parametric entropy. Now when one plots the dimensionality $D_f$ versus 
parametric entropy, $H'_1$, as shown in Fig.2,  a trend appears indicating a
correlation between $D_f$ and $H'_1$, that is between local dynamics and
global dynamics. While no analytical relation could be conjectured for 
$D_f = {\cal F}(H'_1)$, Fig.2 suggests that a statistical analysis performed on 
a larger number of music pieces should provide a better quantification.

 \begin{figure}
\begin{center}
{\resizebox{9cm}{7cm}
{\includegraphics{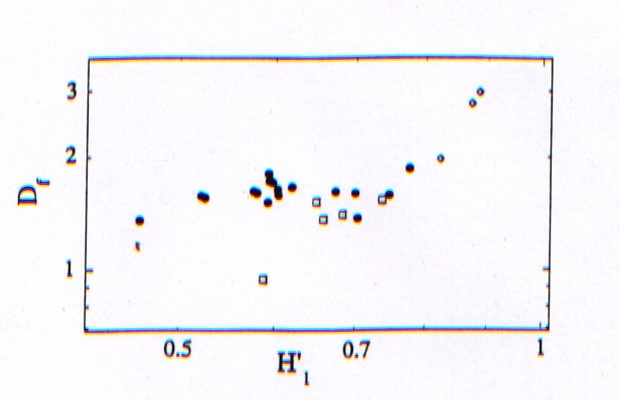}}}
\caption{Dimensionality $D_f$ versus parametric entropy $H'_1$ for about 30
pieces sampled in the music literature from the 17th to the 20th century
(from Boon and Decroly [2]).}
\label{fig:portraits}
\end{center}
\end{figure}

\section*{\large Complexity and artistic value}

 The unexpected elements in a piece of music can be found in the 
deviations from established rules and the violation or even the mere
 rejection of such rules.  In the context of classical forms, these 
deviations are mostly related to the liberty taken by the composer
 with respect to tonality.  Thus when Leibowitz [4]
 considers {\it the complexity of musical language}, 
he argues that {\it Bach's and Haendel's complex polyphonic style is commonly 
opposed to what has been called the homophony of Haydn and Mozart (...). 
 According to which criteria does one evaluate simplicity and complexity? 
 Only one: the counterpoint} (...).  However, continues Leibowitz, {\it the 
counterpoint is hardly the only constituting element in music, and, even
 more, it should be obvious that music can be simple or complex 
independently of any notion of counterpoint}.  Leibowitz then considers 
the problem of harmony and so observes that the composer's {\it audacity as well as 
harmonic complexity may and must be evaluated according to further criteria}.  Those
 then invoked concern {\it  the principles of tonality expansion} and here - as 
argued on the basis of a few specific examples - { \it Haydn's and Mozart's 
works appear more audacious than those of their precursors}.  Obviously 
the argument is of considerable importance as it leads Leibowitz to the 
concept of {\it increasing complexity which should determine the overall 
evolution of musical tradition}.  Considering that entropy provides a 
quantitative measure of the 
degree of complexity the present results show that complexity - in contrast 
with Leibowitz' hypothesis - appears to be characteristic of the composition 
rather than of the composer.  Accordingly we find no indication of 
a systematic increase in complexity paralleling historically the 
evolution of classical music.

What we have considered is how complexity can be identified in music from
the viewpoint of the dynamical nature of the musical object. Obviously there
are aspects of music which have been ignored in this approach, such as the
structural and spectral components of harmony and sound where complexity
is to be identified with other tools. A global characterization of complexity in 
music implies a higher degree of complexity and therefore requires a
sophisticated combination of various complementary approaches. Nevertheless
a striking observation is that in both the analysis of complex structures in 
Pollock's abstract paintings and in the analysis of the dynamics of music pieces 
an identification of the complexity follows from the computation of the 
dimensionality, suggesting that the fractional nature of art might have an
intrinsic value of more general significance.

\bigskip
 
\noindent{\bf References}

\bigskip

\noindent [1] P.W. Anderson, {\em More is different}, Science, {\bf 177}, 393 (1972).

\noindent [2] J.P. Boon and O. Decroly, {\em Dynamical systems theory for
music analysis}, Chaos, {\bf 5}, 501 (1995).

\noindent [3] R.P. Taylor, A.P. Micolich and D. Jonas,
{\em Fractal analysis of Pollock drip paintings}, Nature, {\bf 399}, 422 (1999).

\noindent [4] R. Leibowitz,~{\em L'\'evolution de la musique de Bach \`a
 Sch\"on\-berg}, (Correa, Paris, 1951), Chap.2.

\end{document}